\documentclass[letterpaper]{emulateapj}
\usepackage[colorlinks,urlcolor=blue,citecolor=blue,linkcolor=blue]{hyperref}

\newcommand{\feh}{[\mathit{Fe/H}]} 
 
\newcommand{\mMv}{\mathit{(m-M)_V}}

\newcommand{\EVI}{\mathit{ E(V-I)}}
\newcommand{\EBV}{\mathit{ E(B-V)}}
\shortauthors{Fraga et al.}

\begin{document}

\title{SOAR Adaptive Optics Observations of the Globular Cluster NGC~6496$\dagger$}

\author{Luciano Fraga \altaffilmark{1}}
\affil{Southern Observatory for Astrophysical Research,   Casilla 603, 
La Serena, Chile}
\email{lfraga@ctio.noao.edu}

\and

\author{Andrea Kunder \altaffilmark{2} and Andrei Tokovinin \altaffilmark{2}}
\affil{Cerro Tololo Inter-American Observatory,   Casilla 603, La Serena,
Chile}

\begin{abstract} We present high--quality $BVRI$ photometric data in the field of globular cluster NGC~6496 obtained with the SOAR Telescope Adaptive Module (SAM). Our observations were collected as part of the ongoing SAM commissioning. The distance modulus and cluster color excess as found from the red clump is $\mMv = 15.71 \pm 0.02$\,mag and $\EVI = 0.28 \pm 0.02$\,mag. An age of $10.5 \pm 0.5$\,Gyr is determined from the difference in magnitude between the red clump and the subgiant branch.  These parameters are in excellent agreement with the values derived from isochrone fitting. From the color-magnitude diagram we find a metallicity of $\feh = -0.65$\,dex and hence support a disk classification for NGC~6496. The complete $BVRI$ data set for NGC~6469 is made available in the electronic edition of the Journal.

\end{abstract}


\section{Introduction}

\label{sec:intro} 

The metal-rich globular cluster (GC) NGC~6496 is located at $\alpha_{2000} = 17^{h} 59^{m} 02\fs0$, $\delta_{2000} = -44\degr 15\arcmin 54\arcsec$ and Galactic coordinates of $l=348.02\degr$, $b=10.1\degr$. The compilation by \citet{H96} gives the following distances: $11.3$\,kpc from the Sun, $4.2$\,kpc from the Galactic center and $-2$\,kpc below the Galactic plane.

The first  color-magnitude diagram (CMD) of NGC~6496  was published by \citet{AR88}, reaching  $\sim 2$\,mag below the level  of its horizontal branch (HB).  The cluster harbors a prominent red clump (RC), and its slight tilt was later  interpreted to be  the result  of differential reddening across the cluster  \citep{O99}. Indeed, the reddening maps of \citet{S98}in this region give  a range of  $\EBV$ from  $0.22$ to $0.24$\,mag.

The fundamental parameters that describe NGC~6496 have a considerable range in the literature. \citet{ZW84}  derived a  metallicity of  $\feh = -0.48 \pm 0.15$\,dex and a reddening of $\EBV = 0.09 \pm 0.04$\,mag.  From Washington photometry, \citet{FG91} deduced  a much larger reddening of $\EBV = 0.19$\,mag and a  metallicity of $\feh =-1.05$\,dex, noting that increasing the  reddening to $\EBV  =0.23$\,mag would result  in $\feh= -0.80$\,dex, a metallicity in better  agreement with the estimates based on CMD features.  In a study of six metal--rich GCs, \citet{SN94} present $B$  and $V$  photometry for  the red  giant branch  (RGB) and  the HB region of NGC~6496, finding $\feh =  -0.48$\,dex and $\EBV = 0.22$\,mag from the  cluster's RGB color  and slope.  From new $BV$ photometry, \citet{RG94} use isochrones to  obtain $\feh = -1.03$\,dex, $\EBV = 0.24$\,mag,  assuming a distance modulus $\mu_{0} = 14.82$\,mag and an age of $16$\,Gyr. Recently, \citet{HST03} used Hubble Space Telescope (HST) observations with the F606W and F814W filters to determine a metal content of [M/H]=--0.5, or  $\feh \simeq -0.7$\,dex, $\EBV\ = 0.25$\,mag, a distance modulus of $\mu_{0} = 14.8$\,mag and an age of $10$\,Gyr, from isochrone fitting.

The lack of consensus is not only confined to the cluster's metallicity, extinction, and distance. Its classification as a disk or  halo cluster is also uncertain. Although originally NGC~6496 was believed to be a member of the GC disk system, its inclination was found by \citet{RG94} to be abnormally large for a disk cluster.   A recent comparison of  the shape of  the NGC~6496 and 47~Tuc  main  sequence  (MS)   suggests  that  the  NGC~6496  is  more metal-rich  than  what   \citet{RG94}  adopted,  making  it  extremely unlikely that NGC~6496 is a halo object \citep{HST03}.  Unfortunately, the MS comparison stops before  the main sequence turnoff (MSTO) which is a particularly critical region to constrain theoretical isochrones, because the photometry of 47~Tuc is saturated there.

In this  paper we present  results of $BVRI$ observations  of NGC~6496 using the  SOAR  Adaptive   Module  (SAM)  installed  at  the  4.1\.m SOAR\footnote{$\dagger$ Based   on  observations   obtained   at  the Southern Astrophysical  Research (SOAR) telescope, which  is a joint   project of   the  Minist\'{e}rio  da   Ci\^{e}ncia,  Tecnologia,  e   Inova\c{c}\~{a}o (MCTI)  da Rep\'{u}blica Federativa  do Brasil, the   U.S. National  Optical Astronomy Observatory  (NOAO), the University   of  North  Carolina  at Chapel  Hill  (UNC),  and  Michigan  State   University  (MSU).} telescope  at Cerro  Pach\'{o}n  (Chile).  These observations extend several magnitudes below the MSTO region, allowing new CMDs  to be  constructed and interpreted. 

The next  sections are organized   as  follows:   in  Section   \ref{sec:inst} we  describe SAM.   Sections  \ref{sec:obs}   and   \ref{sec:calib}  describe the observations and  the data reduction procedure.   The color--magnitude diagrams  and derivations  of the  fundamental cluster  parameters are presented  in Section  \ref{sec:cmd}.  Finally, Section  \ref{sec:sum} summarizes our main conclusions.

\section[]{Instrument}
\label{sec:inst}

The SOAR Adaptive Module (SAM) improves image quality by partial correction of turbulence near the ground \citep{SAM10,SAM12}. Ultraviolet light from a pulsed laser scattered by air molecules is used to create the laser guide star (LGS) and to measure wavefront distortion with a 10$\times$10 Shack-Hartmann sensor. The distortion is compensated by a 60-element curvature deformable mirror (DM) inside the re-imaging optics of SAM which projects the uncompensated input focal plane of the telescope to the compensated focus of science instruments, with the same scale. The SAM re-imaging system has substantial distortion, typical of optical relays with two off-axis parabolic mirrors. The UV laser light shortward of 370\,nm is separated from the science beam by a dichroic. Tip-tilt guiding is provided by two probes in the input (uncorrected) focal plane, using stars outside the main field and compensating tilts with the SOAR fast tertiary mirror.

The SAM can feed corrected images either to a visitor instrument \citep[e.g. the high-resolution camera, see][]{HRCam} or to the internal wide-field optical imager, SAMI.  SAMI contains a single CCD with 4096$\times$4112 square pixels of 15\,$\mu$m size from {\it   e2v}. The pixel scale is 45.4\,mas and the total field of view is 186$\arcsec$ across. The CCD is operated with the SDSU-III controller which reads the full unbinned chip in 10\,s with a noise of 3.8 electrons (without patterns) and a gain of 2.1 $e^-/\hbox{ADU}$. SAMI is equipped with standard Bessell $BVRI$ filters and the H$\alpha$ filter with 655.9/64\,nm pass-band.

\section[]{Observations}

\label{sec:obs}

Observations were taken as part of SAM engineering tests, and a log of the observations is given in Table~\ref{tab:Journal}.  The loop was closed and opened several times with the tip-tilt guiding  always on to characterize the resolution gain of SAM.  As the focus in closed loop (CL) was accurately determined, focusing in open loop (OL) was easily done by measuring focus with the LGS.  The flatness of the DM in OL was assured by its calibration which accounts for variable gravity. Therefore, the delivered image quality (DIQ) in OL was not degraded by the SAM optics or defocus. Data was taken without binning to increase the dynamic range of the CCD.

\begin{figure}
\includegraphics[width=\columnwidth]{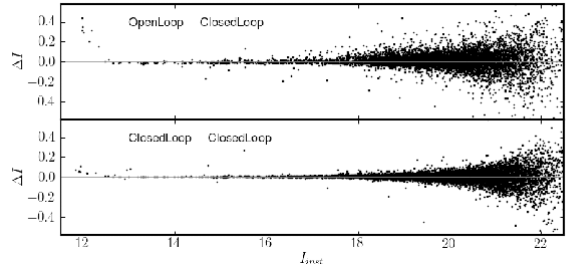}
\caption{The difference in $I$  magnitudes derived by  PSF photometry for common stars found  in two consecutive 120\,s exposures acquired  in open and  closed  loop  (top), and  in  two closed--loop images (bottom). The  horizontal line designates the  zero difference.  Use of SAM greatly increases the photometric accuracy for stars especially in the core of NGC~6496.}
\label{fig:OPvsCL}
\end{figure}

The improvement in the DIQ from the use of SAM depends on the strength of  uncompensated turbulence  in  the upper  atmosphere  (above a  few kilometers).  From the excellent and stable seeing in OL we infer that the atmospheric  turbulence was weak, but likely  variable.  In future observations,  the site  monitor  at Cerro  Pach\'on  will provide  an estimate of  the sky conditions  (this monitor was not  working during our observations).  SAM provided  a substantial improvement of the DIQ in CL, especially  at longer wavelengths. The typical  FWHM of stellar images in CL/OL condition (averaged over the field) was 0.48$\arcsec$/0.59$\arcsec$ in $B$, 0.43$\arcsec$/0.66$\arcsec$ in $V$, 0.32$\arcsec$/0.54$\arcsec$ in $R$, and 0.31$\arcsec$/0.50$\arcsec$  in  $I$. This improvement is further illustrated in Figure~\ref{fig:OPvsCL}. Here PSF photometry  is performed in two  OL and CL images,  as well as in two  CL images. The consistency in the  $I$  magnitudes is worse when introducing OL photometry, likely due to a greater number of unresolved  stellar blends in the  OL image. The photometry of  central stars in NGC~6496 can be obtained to  greater  accuracy  when  using SAM.   Similarly, fainter  magnitudes  can  be  reached  with  a  more  strongly  peaked point-spread function (PSF) in CL, resulting in a deeper photometry at a given exposure time.

The images of NGC~6496 in CL show  a trend of the DIQ in the East-West direction. In  the worst case,  the FWHM in  the $I$ band  varies from $0.36\arcsec$ in the western side of the image to $0.26\arcsec$ in its eastern side.  We believe this  trend originates from a combination of two effects:  guiding with  only one probe  and a delay  in high-order compensation  which  displaces  the  best correction  from  the  field center. The sharpest stars with  the smallest PSF are located closest to the guiding  probe; when guiding with two  probes, the DIQ gradient is less steep.

\begin{table}
\centering
\tabcolsep 0.1truecm
\fontsize{9} {12pt}\selectfont
%\tabcolsep 0.20truecm 
\caption{Log of observations of the globular cluster NGC~6496 and standard star fields 
  observed with SAMI.\label{tab:Journal}} 
\begin{tabular}{lcccc}
  \hline
  \hline
  \noalign{\smallskip}
  Target& Date &Filter& Exposure (sec) & airmass\\
  \noalign{\smallskip}
  \hline
  \noalign{\smallskip}
  NGC~6496      & 2012 May 08    &\textit{B} & 1x120        &1.08\\
  Closed loop  &                &\textit{V} & 2x120, 1x300 &1.07$-$1.14\\
  &                             &\textit{R} & 2x120        &1.07$-$1.09\\
  &                             &\textit{I} & 6x120, 2x300 &1.07$-$1.19\\
\hline
  PG~1633      & 2012 June 06   &\textit{B}& 5x40      &1.75$-$1.78\\
  Open loop&                    &\textit{V} & 5x20      &1.79$-$1.81\\
  &                             &\textit{R} & 5x15      &1.73$-$1.71\\
  &                             &\textit{I} & 5x10      &1.74$-$1.73\\
\hline
  PG~1323      & 2012 June 06   &\textit{B} & 5x30      &1.08\\
  Open loop&                    &\textit{V} & 5x15      &1.08\\
  &                             &\textit{R} & 5x10      &1.08\\
  &                             &\textit{I} & 5x7       &1.08\\
\hline
  SA~107       & 2012 June 06   &\textit{B} & 5x30       &1.33\\
  Open loop&                    &\textit{V} & 5x20       &1.36$-$1.37\\
  &                             &\textit{R} & 5x15       &1.35\\
  &                             &\textit{I} & 5x10       &1.34\\
\noalign{\smallskip}
\hline
\end{tabular}
\end{table}

\begin{figure*}
\centering
  \includegraphics[width=16cm]{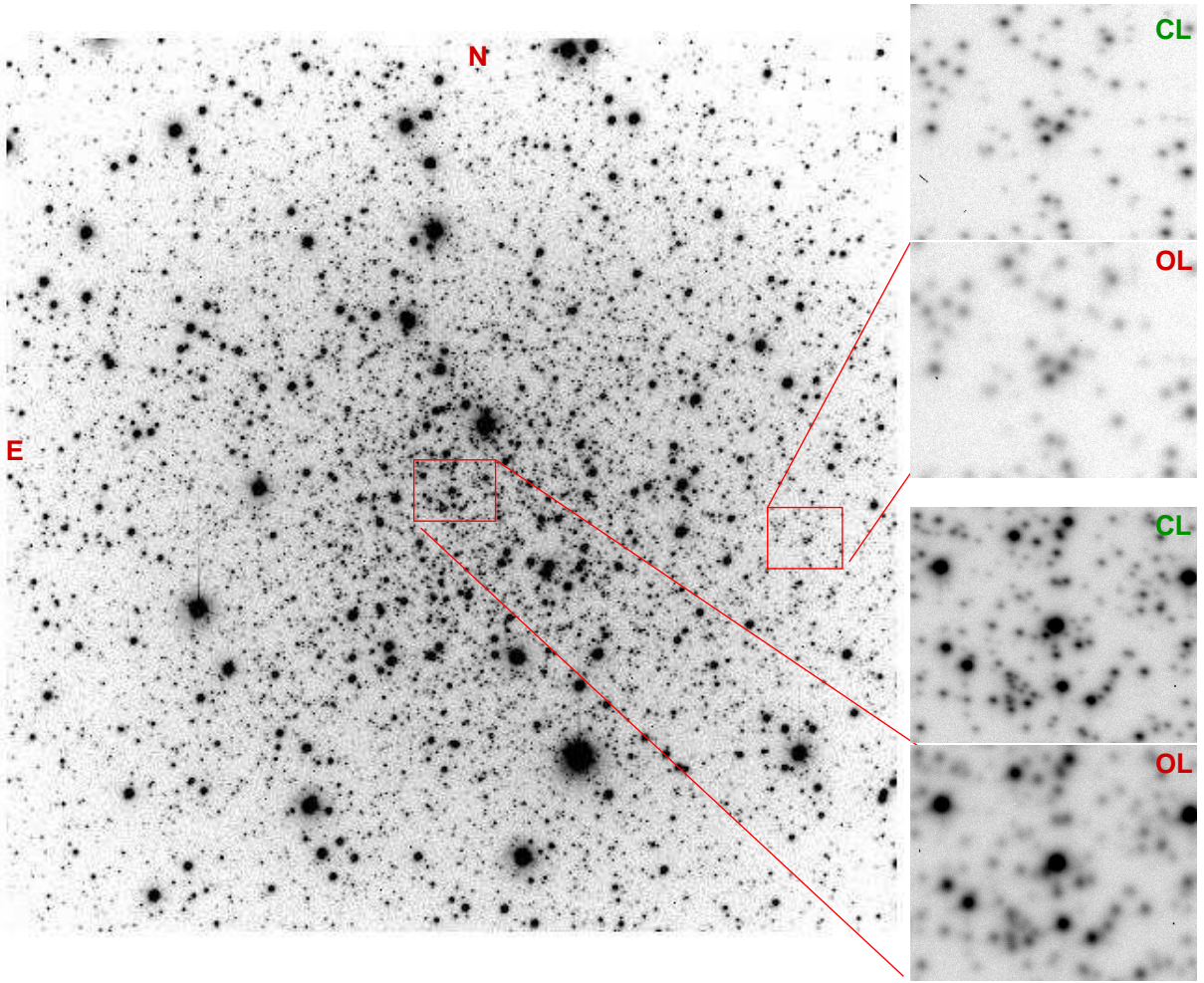}
\caption{Full-frame image of NGC~6496 in the $I$ band taken with SAMI (North  up, East  to the  left).  The  enlarged fragments  of $15\arcsec \times 12\arcsec$ size compare closed-loop (upper) and open-loop (lower) images taken with the same exposure time of 120\,s and displayed on the same  intensity scale, at the  center and near the  edge of the 186\arcsec ~field.}
\end{figure*}

\section{Data Reduction \& Calibrations}
\label{sec:calib}
The CCD data frames were reduced in the standard manner using the mosaic reduction (MSCRED) package in IRAF\footnote{IRAF is distributed by the National Optical Astronomy Observatories, which are operated by the Association of Universities for Research in Astronomy, Inc., under cooperative agreement with the National Science Foundation.} \citep{MSCRED98a,MSCRED98b}. The data reduction process includes bias subtraction, flat-fielding, and cosmic-ray cleaning.

\subsection{Astrometric calibration}
The astrometric solutions were derived using the IRAF task {\bf ccmap} and a tangent-plane projection with a third order polynomial (TNX projection) was used to model the image distortions present in SAMI.  About 150 stars from the 2MASS All-Sky Point Source Catalogue \citep{2MASS03} were used as astrometric reference. The typical astrometric accuracy of an individual SAMI science frame is better than 45 mas (or 1 pixel) in both right ascension and declination.

\subsection{PSF-Fitting Photometry} 

Instrumental magnitudes were determined from PSF-fitting  photometry on individual frames using {\bf daophot~II} and  the task {\bf allstars}  \citep{Stetson87,Stetson92} under IRAF. We selected $\sim$80-100 unsaturated and isolated  bright stars  as PSF-reference stars and used  the Moffat ($\beta = 2.5$) function  to model the PSF. These  same  stars  were  then  also used  to  determine  an  aperture correction for each image.  We define an aperture that encompasses the range  of FWHMs  resulting  from the  trend  in the  DIQ mentioned  in Section  \ref{sec:obs}, and  also  allow the  PSF  parameters to  vary quadratically  across the  field of  view.   In this  way the  typical normalized scatter of the PSF fits is less than 2.5\% .

\subsection{Photometric calibration}

\begin{table*}
\begin{center}
\tabcolsep 0.2truecm
\fontsize{8} {8pt}
%% comments: The full 
\caption{$BVRI$ photometry of  7416 stars in the field of NGC~6496. \label{tab:phot}}\begin{tabular}{ccccccccccccc}
\hline
\hline
\noalign{\smallskip}
ID & RA (J2000) & DEC (J200) & $B$ & $\sigma_{B}$ & $V$ & $\sigma_{V}$ & $\epsilon(V)$ & $R$ & $\sigma_{R}$& $I$ & $\sigma_{I}$& $\epsilon(I)$ \\
\noalign{\smallskip}
&[h~m~s] &[$\degr~\arcmin~\arcsec$] & [mag] & [mag]  &  [mag] &  [mag ] &  [mag] & [mag] &  [mag] &  [mag]  &  [mag] &  [mag] \\
\hline
\noalign{\smallskip}
\vdots&\vdots&\vdots&\vdots&\vdots&\vdots&\vdots&\vdots&\vdots&\vdots&\vdots&\vdots&\vdots\\
1201  &17:58:59.41 &-44:14:51.4 &17.479 & 0.004 &16.405 &0.002 &0.007&15.812 &0.001 &15.216& 0.002&0.009\\
1202  &17:59:02.59 &-44:15:40.0 &17.491 & 0.005 &16.402 &0.003 &0.011&15.794 &0.003 &15.214& 0.004&0.003\\
1203  &17:59:02.72 &-44:15:09.2 &17.492 & 0.004 &16.234 &0.002 &0.006&15.546 &0.002 &14.879& 0.001&0.008\\
1204  &17:59:06.08 &-44:16:46.3 &17.494 & 0.004 &16.390 &0.002 &0.008&15.795 &0.002 &15.204& 0.002&0.008\\
1205  &17:59:07.66 &-44:16:57.9 &17.495 & 0.004 &16.417 &0.002 &0.008&15.832 &0.002 &15.255& 0.002&0.010\\
1206  &17:59:09.49 &-44:16:40.4 &17.500 & 0.004 &16.387 &0.002 &0.007&15.797 &0.001 &15.203& 0.001&0.008\\
1207  &17:59:01.72 &-44:15:14.8 &17.503 & 0.004 &16.431 &0.002 &0.009&15.850 &0.001 &15.254& 0.001&0.008\\
\vdots&\vdots&\vdots&\vdots&\vdots&\vdots&\vdots&\vdots&\vdots&\vdots&\vdots\\
\noalign{\smallskip}
\hline
\end{tabular}
\end{center}
\centering
{\footnotesize Note: The full version of this table is available in
  the electronic edition or from  the authors by request.}
\vspace{10mm}
\end{table*}

In  order to  transform instrumental  magnitudes and  colors  into the standard system,  calibration images were  obtained with SAMI  (in OL) under  photometric conditions  during the  night of  2012 June  6 (see Table~\ref{tab:Journal}).  The zeropoint,  color terms, and extinction were then determined  using  the  IRAF  package {\bf  photcal} through a uniformly  weighted  fit of  the following transformation equations:
\\\ {\small
\\ $b  = B - (0.06\pm0.03) - (0.21\pm0.01)(B-V) + (0.18\pm0.02)\,X$,
\\ $v  = V - (0.48\pm0.01) + (0.01\pm0.01)(V-R) + (0.11\pm0.01)\,X$,
\\ $r  = R - (0.79\pm0.01) - (0.03\pm0.01)(V-R) + (0.04\pm0.01)\,X$,
\\ $i  = I - (0.16\pm0.01) - (0.07\pm0.01)(R-I) + (0.02\pm0.01)\,X$,
\\ }
\\
where $B$, $V$, $R$, and $I$ are the magnitudes in the standard system,  $b$,  $v$, $r$,  and  $i$  are  the instrumental  magnitudes, corrected  to  the  exposure  time  of  1\,s using  a  zero  point  of 25.0\,mag, and  $X$ is the air-mass at the time of the observations. The  r.m.s. deviations from the fits  were 0.013, 0.013, 0.010 and 0.018 mag in $B,V,R$, and $I$, respectively.

We used the above photometric calibration to transform the NGC~6496 instrumental magnitudes into the standard system for the $VRI$ passbands.  However, given that the standard stars were not observed on the same night as our observations of NGC 6496, the zero points were adjusted from the Stetson (2000) Standard Fields found in the images \footnote{http://www3.cadc-ccda.hia-iha.nrc-cnrc.gc.ca/community/}.  In the $V,R$ passbands, eight Stetson standards are in common with stars in our field,  and seven stars are in common with Stetson standards in the $I$ passband. The mean difference, in the sense of this work minus that of \citet{Stetson00}, are as follows: $\langle\Delta V\rangle = 0.038 \pm 0.01$, $\langle\Delta R\rangle = -0.004 \pm 0.018$, $\langle\Delta I\rangle = 0.038 \pm 0.02$,$\langle\Delta (V - R)\rangle = 0.042 \pm 0.012$, $\langle\Delta (V - I)\rangle = 0.000 \pm 0.012$, $\langle\Delta (R - I)\rangle = 0.040 \pm 0.013$.  In this way, the above values were used to adjust the zero points in $VRI$ filters. This accounts for potential small differences in the extinction coefficients between our observations of object and standards. Apart from the zero points differences, no significant color trends was found. Since \citet{Stetson00} do not include $B$ photometry in the field of NGC~6496, our $B$ zero point might be off by few per cent.  

Our photometry as obtained from SAM and transformed to the standard system described above is presented in Table~\ref{tab:phot}. Our observations include 7416 stars in the $BVRI$ passbands. The uncertainties $\sigma_\lambda$ in the $BVRI$ passbands are formal errors from the PSF fitting, while the uncertainties $\epsilon(\lambda)$ in the $VI$ passbands are the rms frame-to-frame variations. The formal errors from the PSF fitting and the frame-to-frame differ by a factor 2-3, but indicate that for the most part, our photometry is accurate at the 1\% level until $V\sim$19.0 mag. The errors listed do not include those from the standard star transformations.

\section{Analysis and Results} 
\label{sec:cmd}

\begin{figure*}
\includegraphics[width=18cm]{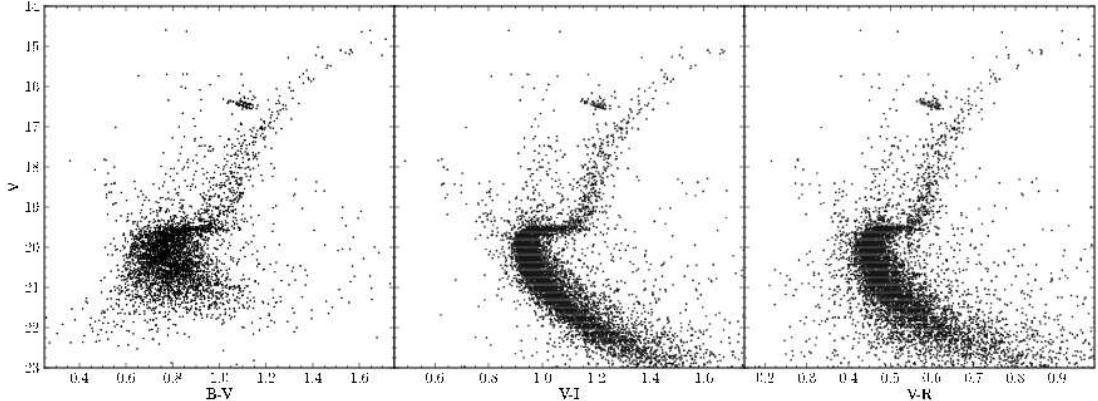}
\caption{The [$B-V,V$], [$V-I,V$] and [$V-R,V$] color-magnitude diagrams of the globular cluster NGC~6496 for a total of 4168, 7416 and 7373 stars, respectively, in a $186\arcsec\times186\arcsec$ field approximately centered on the cluster.}
\label{fig:cmdall}
\end{figure*}

\begin{figure}
\plotone{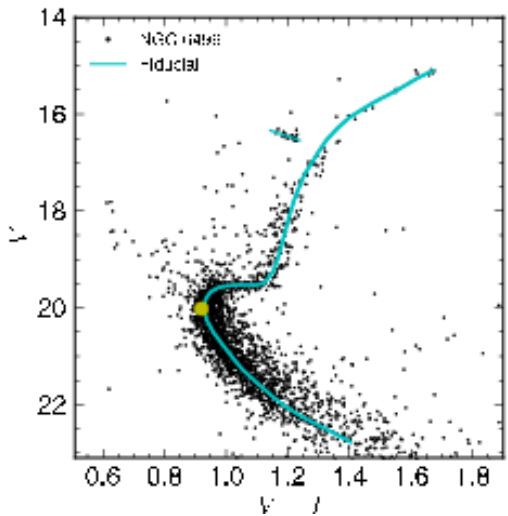}
\caption{The [$V-I,V$] color-magnitude diagram of NGC~6496 for the stars within 90\arcsec of the cluster center. The solid line (cyan) represents the mean ridge-line tracing the evolutionary features of the cluster, and the circle (yellow) represents the main sequence turnoff point.}
\label{fig:fid}
\end{figure}

\begin{figure}
\centering
\includegraphics[scale=0.8]{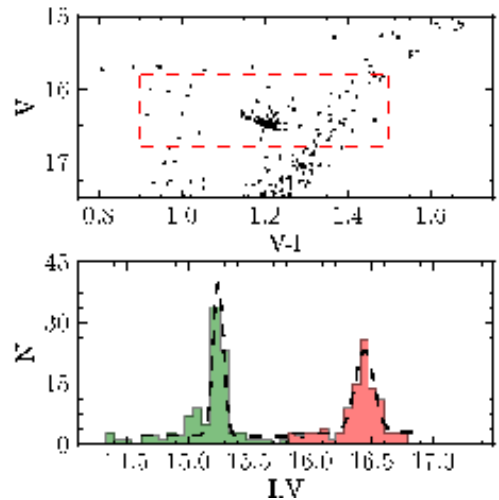}
\caption{The top panel shows a zoom-in of the red clump (RC) region in the [$V-I,V$] CMD of NGC~6496. The dashed rectangle (red) encloses 103 stars selected to estimate the apparent magnitude of the RC. The bottom panel shows the $V$ (red) and $I$ (green) apparent magnitude distribution of those stars. Over-plotted are the results of a non-linear least-squares fit, given in Equation~\ref{eq:rc}}
\label{fig:rc}
\end{figure}

\begin{figure*}
\centering
\plottwo{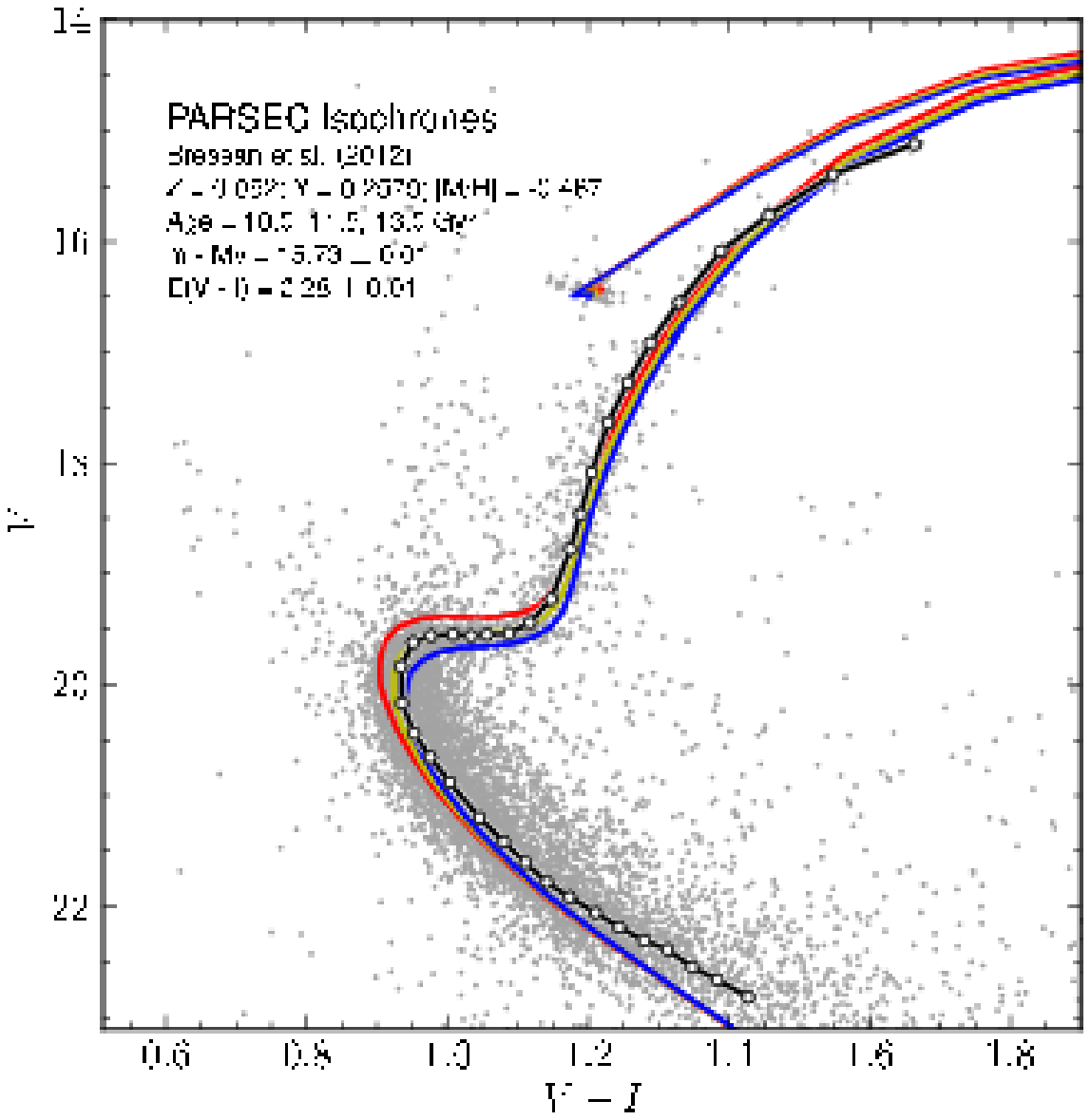}{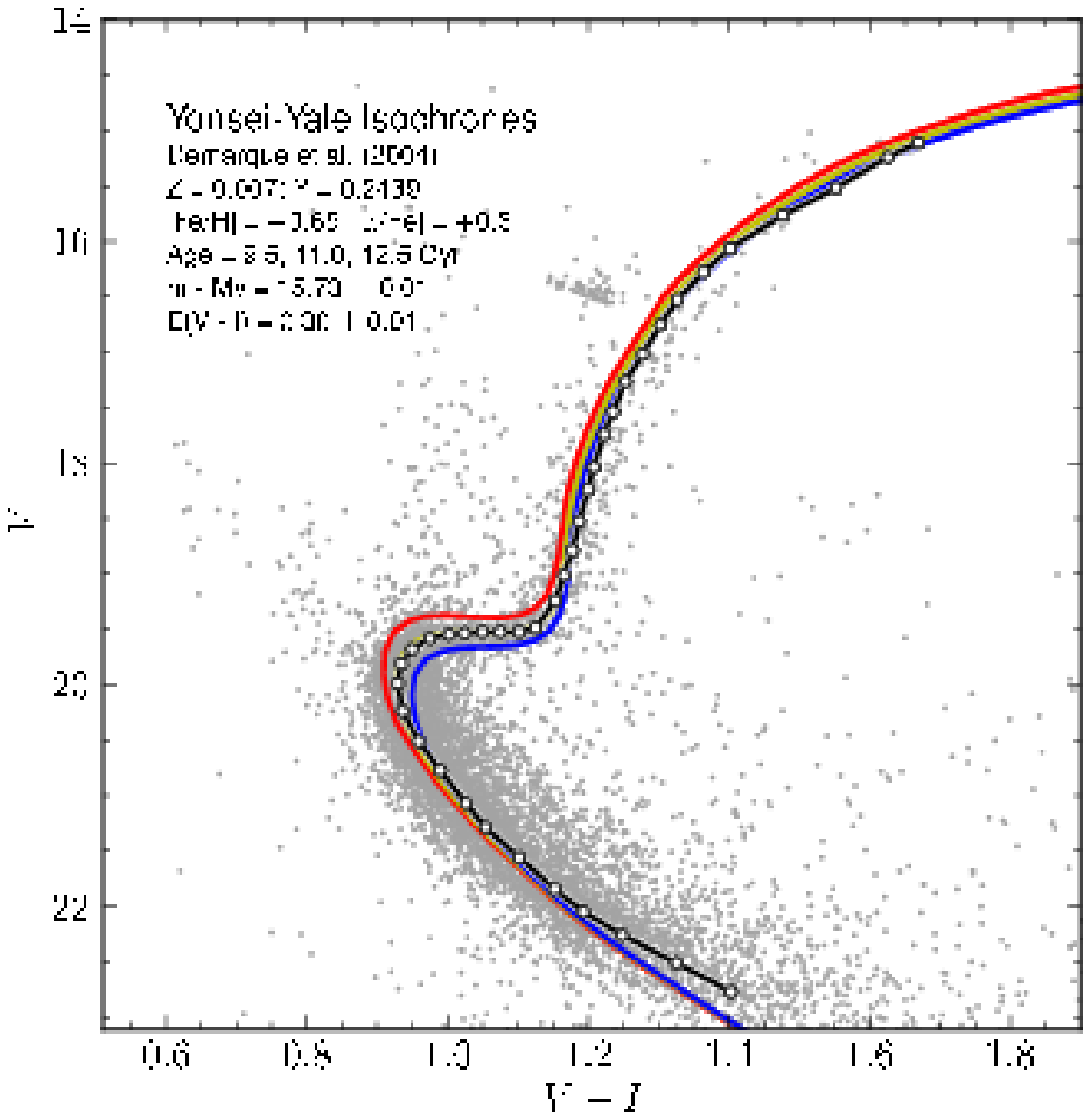}
\caption{{\it Left:} The observed [$V-I,V$] CMD of NGC~6496 with superimposed PARSEC isochrones covering an age range of 10.0, 11.5 and 13 Gyr. The fit parameters are listed in the plot. The fiducial points are represented by the circles. {\it Right:} Same as the left panel, but for the Yonsei--Yale isochrones covering an age range of 9.5, 11.0, 12.5 Gyr.}
\label{fig:iso}
\end{figure*}

\subsection{Color-Magnitude Diagram}

Figure \ref{fig:cmdall} presents the CMDs with three different color baselines in a $186\arcsec \times 186\arcsec$ field approximately centered on NGC~6496. The deepest exposures were obtained in the $VI$ filters, for that reason, the subsequent analysis will focus on the photometry data from those filters. Figure~\ref{fig:fid} shows the [$V-I,V$] CMD for all stars within a radius of 90$\arcsec$ from the center  of NGC~6496. The location of the cluster center on the CCD was estimated from the peak position of the distribution of stars in the x- and y-directions. The (x,y)-coordinates were transformed to right ascension and declination and found to be $\alpha_{2000} = 17^{h} 59^{m} 03\fs5$, $\delta_{2000} = -44\degr 15\arcmin 57.9\arcsec$, in excellent agreement with $\alpha_{2000} = 17^{h} 59^{m} 03\fs7$, $\delta_{2000}=-44\degr 15\arcmin 57.4\arcsec$ published in \cite{G10}.

The solid line (cyan) in the Figure~\ref{fig:fid} represents the adopted mean ridge line tracing the evolutionary features of the cluster. The ridge line has been derived separately for the MS, SGB and RGB. For the MS and the RGB, a set of preliminary fiducial points separated by 0.25 mag in $V$ was determined by eye and then refined by averaging the color distribution of all stars with $(V-I)$ color less than 0.025 mag. Similarly, for the SGB region ($19.4 < V < 19.7$ mag) the preliminary points were spaced by 0.025 mag in $V-I$ and then refined by averaging the $V$ magnitude distribution of all stars with $V <$ 0.25 mag from the preliminary points. A linear fit was used to determine the HB fiducial line. 

The apparent magnitude of the MSTO, the bluest point along the MS, was estimated by fitting a third order polynomial to the mean ridge line in the range of $19.5 < V < 21$ mag. This magnitude range encompasses a generous area with which to determine the MSTO (see Figure~\ref{fig:fid}). This yielded a turnoff apparent magnitude of $V_{TO} = 20.02 \pm 0.04$ and color of $(V-I)_{TO} = 0.92 \pm 0.02$, marked with a circle (yellow) in the same figure. This value is slightly brighter than the  $V_{TO} = 20.15 \pm 0.05$ reported by \cite{RG94} and subsequently supported by \cite{HST03}. The large photometric errors in the \cite{RG94} CMD below the turn-off may be one reason for our discrepancy. 

The apparent $V$ magnitude of the SGB, $V_{SGB}$, is obtained by using the mean magnitude of the stars in a box centered at $(V-I)_{TO} + 0.05$ mag, as discussed by \citep{Chaboyer96}. Our adopted box size is 0.04 mag in $(V-I)$ and 0.15 mag in $V$; these parameters were determined from the above mentioned polynomial fit and encompasses the SGB well.  A $V_{SGB} = 19.61 \pm 0.02$ is obtained, and will be used later (in \S5.2) as an age indicator.

The horizontal branch "red clump" \citep[RC,][]{cannon70} is clearly distinguished as a clump of red stars near the RGB. Red clumps are common in CMDs of intermediate-age ($\sim$1--10 Gyr) or very old ($>$10 Gyr) clusters with metal-rich stellar populations \citep[e.g.,][]{alves99}. The top panel of Figure~\ref{fig:rc} displays a zoom-in of the RC region of NGC~6496. The dashed (red) rectangle spanning $15.8 < V < 16.8$ and $0.9 < (V-I) < 1.5$ encloses 103 stars selected to estimate the apparent magnitude of the RC. The histograms in the bottom panel in Figure~\ref{fig:rc} show the $V$ and $I$ apparent magnitude distribution of those stars. The dashed lines represent the result of a non-linear least-squares fit of the function
\begin{equation}   
N(m_\lambda) = a + b m_\lambda + c m_\lambda^2 + d \exp\left[-\frac{(m_\lambda^{\rm RC}-m_\lambda)^2} {2\sigma_{m_\lambda}^2}\right]   
\label{eq:rc}   
\end{equation}   
  to each distribution, as suggested by \cite{SG98}. The apparent magnitude of the RC and its associated standard error in both $V$ and $I$ photometric bands were determined to be $V_{RC} = 16.44 \pm 0.07$ and $I_{RC} = 15.24 \pm 0.03$. The $V_{RC}$ value is identical to that derived by \citet[$V_{RC} = 16.44$]{SN94}, and it is between the values reported by \citet[$V_{RC} = 16.54$]{RG94} and \citet[$V_{RC} = 16.36$]{AR88}, respectively.

\subsection{Age, distance and reddening from the RC}

We estimate here the age of NGC~6496 using the difference in magnitude between $V_{SGB}$ and $V_{RC}$ , $\Delta V_{(SGB - RC)}$ = $V_{SGB} - V_{RC}$ \citep{Chaboyer96, Chaboyer00}. This method is also known as the 'vertical method' and relies on the fact that the absolute brightness of the RC ($M_{V}^{RC}$) remains nearly constant for ages older than 9\,Gyr \citep{GS01}, while the absolute magnitude of the SGB depends on the age of the cluster. The absolute magnitude of the SGB as a function of age was obtained for two values of metallicity ($\feh = -0.49$ and $\feh = -0.65$, adopting [$\alpha/Fe] = +0.3$) using the Yonsei--Yale \citep[Y$^2$]{Y2} isochrones. The adopted value for the $M_{V}^{RC} = 0.73$ was taken from \cite{Alves02}, who has estimated the absolute magnitude of the red clump stars for the local stellar population with Hipparcos parallaxes. Population corrections from \cite{GS01} indicate that $M_{V}^{RC}$ would be negligible for an age and metal content representative of NGC~6496 with the $M_{V}^{RC}$ being affected by 0.02 mag at the most (e.g. assuming an age of $\sim$ 10.5 Gyr and Z=0.004). Therefore the observed value for $\Delta V_{(SGB - RC)} = 3.17 \pm 0.04$, which  corresponds to an age of $10.5 \pm 0.5$\,Gyr. This  is in excellent agreement with the age reported in \citet[][age  = 10\,Gyr]{HST03}.

We next obtain the distance and reddening to NGC~6496 following the method described in \cite{Salaris12}. Using an absolute magnitude for the RC of $M_{V}^{RC} = 0.73$ and the intrinsic RC color of $(V-I)_{RC} = 0.92$, the color excess $\EVI$ was estimated to be $\EVI = 0.28 \pm 0.02$\,mag. Adopting $\EVI/\EBV = 1.28$, $A_{V}/\EBV=3.24$ and $A_{I}/A_{V} = 0.601$ \citep{S98}, the following quantities are derived: $\EBV = 0.22 \pm 0.02$, $A_{V} = 0.71 \pm 0.02$, and $A_{I} = 0.43 \pm 0.02$. The true distance modulus has been evaluated using the equation $\mu_0 = m_{\lambda} - M_{\lambda} - A_{\lambda}$, for both $V$ and $I$ photometric bands.  We find $\mu_0 = 15.0 \pm 0.03$ mag, which corresponds to a  distance of $d_\sun = 10.0 \pm 0.1$\,kpc. The reddening and the distance derived above agree, within the errors, with the published values from \citet[][$\EBV  = 0.22$]{SN94}, \citet[][$\EBV = 0.24$ and $\mu_{0} = 14.82$]{RG94} and \citet[][$\EBV = 0.25$ and $\mu_{0} = 14.8$]{HST03}. Table \ref{tab:obspar} shows a complete list of the parameters determined from our RC analysis.

\begin{table}
\centering
\tabcolsep 0.1truecm
\fontsize{8} {8pt} 
\selectfont
\caption{Results from the RC analysis\label{tab:obspar}} 
\begin{tabular}{lrrr}
\hline
\hline
\noalign{\smallskip}
Parameter & &Value&\\
\noalign{\smallskip}
\hline
\noalign{\smallskip}
$V_{RC}$\hfil&16.44&$\pm$ 0.07&mag\\
$I_{RC}$\hfil&15.24&$\pm$ 0.03&mag\\
$V_{TO}$\hfil&20.02&$\pm$ 0.04&mag\\
$(V-I)_{TO}$\hfil&0.92& $\pm$ 0.02&mag\\
$V_{SGB}$\hfil&19.61&$\pm$ 0.02&mag\\
$(m-M)_v$\hfil&15.71&$\pm$ 0.02&mag\\
$E(V-I)$\hfil&0.28&$\pm$ 0.02&mag\\
$Age$\hfil&10.5& $\pm$ 0.5&Gyr\\
$\mu_0$\hfil&15.0&$\pm$ 0.03&mag\\
$d_{\sun}$\hfil&10.0&$\pm$ 0.1&kpc\\
\noalign{\smallskip}
\hline
\end{tabular}
\vspace{10mm}
\end{table}

\subsection{Isochrone Fitting}

Theoretical isochrones from PARSEC (PAdova and TRieste Stellar Evolution Code, \citealt{PARSEC}) and from Y$^2$ (Yonsei--Yale, \citealt{Y2}) were fitted to the observed [$V - I, V$] CMD to further investigate the reddening, distance, metallicity, and the age of NGC~6496. As a preliminary step, we adopt the distance and reddening determined in the previous subsection (see Table~\ref{tab:obspar}). Next the best-fit isochrone was found by slightly varying the distance, reddening, metallicity and age. The left panel of Figure~\ref{fig:iso} shows the best-fit to the PARSEC isochrones obtained with an age of $11.5 \pm 0.5$\,Gyr, an apparent distance modulus of $15.73 \pm 0.02$, a color excess of $\EVI = 0.28 \pm 0.01$ and a metal content of $[M/H] = -0.47$ ($\feh = -0.67$ assuming $[\alpha/Fe] = +0.3$,~\citealt{Paust10}). The right panel in Figure~\ref{fig:iso} displays the best-fit to the Y$^2$ isochrones using an age of $11.0 \pm 0.5$ Gyr, an apparent distance modulus of $15.73 \pm 0.01$, a color excess of $\EVI = 0.30 \pm 0.01$ and a metallicity of $\feh = -0.65$ ($[\alpha/Fe] = + 0.3$).The distance  ($d_\sun$) derived from the PARSEC and Y$^2$ isochrones fitting,  adopting $\EVI/\EBV = 1.28$ and $A_{V}/\EBV = 3.24$, were determined to be $d_\sun = 10.1 \pm 0.1$\,kpc and $9.8 \pm 0.1$\,kpc, respectively. The above values are in general consistent with the values derived from the RC analysis. Table \ref{tab:isofit} summarizes our results from the isochrone fitting.

Although the best fit to the RGB from the PARSEC isochrones matches the lower RGB and upper RGB well, the middle part of the RGB is not matched as well. We were unable to find a value for Z that better represented the RGB for NGC~6496, however.  In contrast, the best fit Y$^2$ isochrones matches the RGB well. The isochrones also do not match the lower MS very well.  It has been observed previously that the theoretical isochrones are fainter in the lower MS than observed \citep[e.g.,][]{Yi01,An09}. For the Y$^2$ isochrones, for example, \citet{Y2} acknowledge that their isochrones may not be very impressive for fitting the lower MS CMDs.  One reason for this may be because in low-mass MS stars, one finds a convective envelope and, thus, uncertainties about superadiabatic convection \citep[see e.g.][]{castellani99}.

\begin{table}
\centering
\tabcolsep 0.2truecm
\fontsize{8} {8pt} 
\selectfont
\caption{Results from the isochrones fitting\label{tab:isofit}} 
\begin{tabular}{lrr}
\hline
\hline
\noalign{\smallskip}
&\multicolumn{2}{c}{Models} \\
\cline{2-3}
\noalign{\smallskip}
Parameter & PARSEC & Yonsei--Yale\\
\hline
$(m-M)_v$  \hfil&15.73 $\pm$ 0.02 mag &15.73 $\pm$ 0.02 mag\\
$E(V-I)$\hfil&0.28 $\pm$ 0.01 mag  &0.30 $\pm$ 0.01 mag\\
$Age$   \hfil&11.5 $\pm$ 0.5 Gyr   &11.0 $\pm$ 0.5 Gyr\\
$\feh$  \hfil&$-$0.67 $\pm$ 0.05 dex&$-$0.65 $\pm$ 0.05 dex\\
$\mu_0$   \hfil&15.02 $\pm$ 0.02 mag   &14.97 $\pm$ 0.02 mag\\
$d_{\sun}$  \hfil&10.1 $\pm$ 0.1 kpc&9.8 $\pm$ 0.1 kpc\\
\hline
\end{tabular}
\vspace{10mm}
\end{table}

\subsection{Comparison with other GCs}

\begin{figure} 
\centering
\includegraphics[width=9cm]{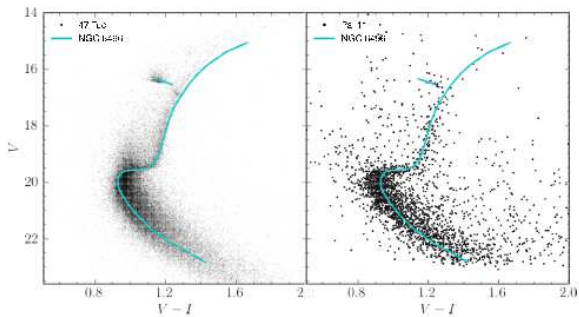}
\caption{Comparison between the fiducial lines of NGC~6496 (solid cyan lines) and the CMDs of 47~Tuc (left panel) and Pal~11 (right panel).}
\label{fig:comp}
\end{figure}

47 Tucanae and Palomar 11 are two metal-rich GCs that can be compared with NGC~6496. The photometric data sets of 47~Tuc and Pal~11 used here are taken from \cite{47Tuc} and \cite{Pal11}, respectively. The fundamental parameters of 47~Tuc listed by \cite{47Tuc} are: age = 12 Gyr, $\feh = -0.83$, $(m-M)_v = 13.37$ and $\EVI = 0.05$. While \cite{Pal11} reported the following values for Pal~11: age = 10.5 Gyr, $\feh = -0.76$, $(m-M)_v = 16.77$ and $\EVI = 0.40$. Figure~\ref{fig:comp} shows the [$V-I,V$] CMDs of 47~Tuc and Pal~11 along with the fiducial ridge-line of NGC~6496 (solid cyan line) superimposed on both CMDs.  For the comparison, the CMD of 47~Tuc is shifted in color by $\Delta (V-I) = 0.25$ and in $V$ magnitude by $\Delta V = 2.38$ to match the fiducial lines of NGC~6496. Similarly, the Pal~11 CMD is shifted by $\Delta (V-I) = -0.11$ and $\Delta V = -1.01$. The applied shifts are in good agreement, within the errors, with the values determined for the apparent distance modulus and color of NGC~6496. For both clusters the SGBs and the RCs overlap remarkably well with the NGC~6496 fiducial lines, suggesting these cluster are similar in age. The Pal 11 RGB overlaps remarkably well with the RGB of NGC~6496, suggesting that both clusters have similar metallicities.  On the other hand, 47 Tuc has a RGB that is slightly bluer than that of NGC~6496. This suggests that NGC~6496 is more metal-rich than 47 Tuc, which again is supported by our isochrone fitting.

Lastly, Pal~11 and the thick-disk GC NGC~5927 have similar CMDs \citep{Pal11}, and hence NGC~5927 is similar to NGC~6496 as well. Unlike Pal~11, NGC~5927 has a spectroscopic Fe measurement based on Fe II lies of $\feh = -0.67$ dex \citep{KI03}, and we note that this metallicity is in excellent agreement with the $\feh$ derived for NGC~6496 from theoretical isochrones (see Table~\ref{tab:isofit}). Therefore an intermediate metal content for this cluster, as suggested by \citet{RG94}, is inconsistent with our observational and theoretical results. We find no need to classify NGC~6496 as a halo object and support its classification as a disk cluster, in agreement with \citet{HST03}.

\section{Concluding remarks}
\label{sec:sum}

The SOAR 4.1m Telescope Adaptive Module (SAM) has been used to obtain $BVRI$ photometric data in the field of NGC~6496. The main results of this paper can be summarized as follows:

\begin{enumerate}

\item The distance modulus of $15.71 \pm 0.02$ mag and a reddening of $\EVI = 0.28 \pm 0.02$ mag has been estimated using the apparent magnitude of the RC and following the method described by \citet{Salaris12}. After correcting for extinction, a true distance modulus of $15.0 \pm 0.02$ mag has been derived, which corresponds to a distance of $10.0 \pm 0.1$ kpc.

\item The age of NGC~6496 has been estimated to be $10.5 \pm 0.5$ Gyr, using the magnitude difference between $V_{SGB}$ and $V_{RC}$ together with the models of Yonsei--Yale and the absolute magnitude of the red clump stars for the local stellar population taken from \cite{Alves02}.

\item Theoretical isochrones from PARSEC and from Yonsei--Yale were fitted to the observed [$V - I, V$] CMD of NGC~6496. The resulting parameters derived from the isochrone fits (see Table~\ref{tab:isofit}) are in excellent agreement with our results from the RC analysis. The metallicity estimates from both models are very similar, $\feh \sim -0.65$ dex (adopting $[\alpha/Fe] = +0.3$). 

\item A comparison of the NGC~6496 CMD with 47~Tuc and Pal~11 CMDs suggests the classification of NGC~6496 as a disk cluster, in agreement with \citet{HST03}.

\end{enumerate}

\acknowledgments

The Southern Astrophysical  Research (SOAR) telescope is a joint  project of  the  Minist\'{e}rio  da   Ci\^{e}ncia,  Tecnologia,  e   Inova\c{c}\~{a}o (MCTI)  da Rep\'{u}blica Federativa  do Brasil, the   U.S. National  Optical Astronomy Observatory  (NOAO), the University   of  North  Carolina  at Chapel  Hill  (UNC),  and  Michigan  State   University  (MSU). The authors are grateful to the SAM team and SOAR staff for providing an excellent technical support before and during the observations. We thank Peter Bergbusch and Peter Stetson for providing us with the 47~Tuc data.

%\clearpage

\end{document}